# Two-Component Axionic Dark Matter Halos

Gennady P. Berman[1], Vyacheslav N. Gorshkov[2], Vladimir I. Tsifrinovich[3], Marco Merkli[4], and Vladimir V. Tereshchuk[2]

[1]Theoretical Division, T-4, Los Alamos National Laboratory, Los Alamos, NM 87545, USA
[2]National Technical University of Ukraine "Igor Sikorsky Kyiv Polytechnic Institute", Kyiv, 03056, Ukraine
[3]Department of Applied Physics, NYU Tandon School of Engineering, Brooklyn, NY 11201, USA
[4]Department of Mathematics and Statistics, Memorial University of Newfoundland, St John's, NL, A1C 5S7, Canada

**Abstract**

We consider a two-component dark matter halo (DMH) of a galaxy containing ultra-light axions (ULA) of different mass. The DMH is described as a Bose-Einstein condensate (BEC) in its ground state. In the mean-field (MF) limit we have derived the integro-differential equations for the spherically symmetrical wave functions of the two DMH components. We studied numerically the radial distribution of the mass density of ULA and constructed the parameters which could be used to distinguish between the two- and one-component DMH. We also discuss an interesting connection between the BEC MF ground state of a one-component DMH and Hawking temperature and entropy of a corresponding Black Hole, and Unruh temperature.

## I.    Introduction

The content of the dark matter in the galaxy halos (DMH) is one of the most important unsolved problems in contemporary physics (see, for example, [1]). One of the possible avenues for description of DMH is a two-component DMH (see, for example [2,3]). In recent years the ultra-light axions (ULA) of mass less than $10^{-17}$ eV, originated in the string theory, have emerged as one of the most promising candidates for the DMH [4-7]. In this paper we combine these two approaches: we consider DMH containing two ULA components of different mass. Following our previous work [7], we assume that every component can be described as the ground state of the Bose-Einstein condensate (BEC).

Our work has three main objectives: 1. Derivation of the integro-differential equations describing 2-component ULA DMH in the mean-field (MF) approximation. 2. Numerical analysis of the radial distribution of the mass density of ULA. 3. Finding the appropriate parameters which allow one to distinguish a 2-component DMH from a one-component DMH. In the last section of our work we demonstrate an intriguing correlation between a single component DMH in the MF approximation and the Black Hole temperature and entropy.
Our main results include:
  1. For a two-component quantum system of non-relativistic gravitationally interacting axions, the coupled nonlinear integro-differential equations, of the Hartree-Fock type, are derived, for the single-particle wave functions, $\phi_{1,2}(\vec{x},t)$, in the mean-field limit: $N_{1,2} \to \infty$, $M_{1,2} = m_{1,2} N_{1,2} = \text{const}_{1,2}$, where $N_{1,2}$, $m_{1,2}$, and $M_{1,2}$ are the number of axions, individual mass of axion, and the total axionic mass of the corresponding component, $\vec{x} = \vec{r}/r_0$, $r_0$ is the characteristic spatial scale of the system. It is demonstrated that the effective potentials include two independent dimensionless parameters, $\alpha = (m_2/m_1)^2$ and $\beta = M_2/M_1$.

2. The numerical algorithm for finding the spherically symmetric BEC ground state, $\phi_{1,2}(x)$, is developed, where $x = |\vec{r}/r_0|$.

3. The single-particle density functions, $p_{1,2}(x) = |\phi_{1,2}(x)|^2$ and the distribution function of axionic mass in the halo, $P(x) = (M_1/M)p_1(x) + (M_2/M)p_2(x)$, are calculated numerically, where $M_1 + M_2 = M$.

4. A parameter, $\hat{\chi}$, is introduced, which allows to distinguish the behavior of $P(x)$ for a two-component axionic system from that of a one-component axionic system.

5. The analogues of the Black Hole Hawking temperature and entropy, and Unruh temperature, are presented for a one-component system.

In the second section of our work, we derive the system of non-linear integro-differential equations which describes the ground state of the two-component BEC, in the MF limit. In the third section, we describe our computer algorithm. In the fourth section, we present the results of numerical computations. In the fifth section, we discuss possible relations between a one-component DMH, in the BEC MF limit, and the corresponding Black Hole temperature and entropy.

**II. Non-Linear Integro-Differential Equations for the Two-Component DMH.**

We start with the non-relativistic Schrödinger equation,

$$i\hbar \partial \psi(\vec{r}_1,\ldots,\vec{r}_{N_1};\vec{r}_1,\ldots\vec{r}_{N_2};t)/\partial t = H_{N_1,N_2} \psi(\vec{r}_1,\ldots,\vec{r}_{N_1};\vec{r}_1,\ldots\vec{r}_{N_2};t), \qquad (2.1)$$

for $N_1 + N_2$ gravitationally interacting axions, described by the Hamiltonian,

$$H_{N_1,N_2} = -\frac{\hbar^2}{2m_1}\sum_{j=1}^{N_1}\Delta_{\vec{r}_j} - \frac{\hbar^2}{2m_2}\sum_{k=1}^{N_2}\Delta_{\vec{r}_k} - Gm_1^2 \sum_{1\le l<s\le N_1}\frac{1}{|\vec{r}_l - \vec{r}_s|} - Gm_2^2 \sum_{1\le p<q\le N_2}\frac{1}{|\vec{r}_p - \vec{r}_q|} - Gm_1 m_2 \sum_{l=1}^{N_1}\sum_{m=1}^{N_2}\frac{1}{|\vec{r}_l - \vec{r}_m|}, \qquad (2.2)$$

where $m_1$, $N_1$, $\vec{r}_i$, are the axion mass, the number of particles and coordinate of the first component, and $m_2$, $N_2$, $\vec{r}_i$, are the axion mass, the number of particles and coordinate of the second component $(\vec{r}_i, \vec{r}_j \in R^3)$, $G$ is the gravitational constant. Introduce the dimensionless coordinates, $\bar{x}_i$ and $\bar{y}_j$: $\bar{x}_i = \vec{r}_i/r_0$, $\bar{y}_j = \vec{r}_j/r_0$, where the dimensional parameter, $r_0$, will be defined below. We have from (2.2),

$$H_{N_1 N_2} \equiv \frac{H_{N_1, N_2}}{\hbar^2/m_1 r_0^2} = -\frac{1}{2}\sum_{j=1}^{N_1} \Delta_{\bar{x}_j} - \frac{1}{2(m_2/m_1)}\sum_{k=1}^{N_2} \Delta_{\bar{y}_k} - \frac{\lambda_1}{N_1}\sum_{1 \le r < s \le N_1} \frac{1}{|\vec{x}_r - \vec{x}_s|} - \frac{\lambda_2}{N_2}\sum_{1 \le p < q \le N_2} \frac{1}{|\vec{y}_p - \vec{y}_q|} - \frac{\lambda_3}{(N_1+N_2)}\sum_{l=1}^{N_1}\sum_{m=1}^{N_2}\frac{1}{|\vec{x}_l - \vec{y}_m|},$$

(2.3)

where the dimensionless constants are introduced,

$$\lambda_1 = \frac{Gm_1^3 r_0 N_1}{\hbar^2},\ \lambda_2 = \frac{Gm_2^2 m_1 r_0 N_2}{\hbar^2},\ \lambda_3 = \frac{Gm_1^2 m_2 r_0 (N_1 + N_2)}{\hbar^2}.$$

(2.4)

We define the characteristic dimensional parameter, $r_0$, from the condition, $\lambda_1 = 1$. This means that $r_0$ is related to the first component. We have from (2.4),

$$r_0 = \frac{\hbar^2}{Gm_1^3 N_1},\ \lambda_2 = \frac{N_2 m_2^2}{N_1 m_1^2},\ \lambda_3 = \frac{(N_1+N_2)m_2}{N_1 m_1}.$$

(2.5)

*MF limit.*

The dimensionless Hamiltonian, $H_{N_1 N_2}$ in (2.3) has a similar form as the dimensionless Hamiltonian, $H_N$ (1.16) (see also (3.7)) in [8], where it was shown that the MF limit can be used when $N_{1,2} \to \infty$. For $N_1 + N_2$ interacting axions, we can consider $N_{1,2}$ very large and total masses of both components, $M_{1,2} = m_{1,2} N_{1,2}$, fixed (as the axion masses, $m_{1,2}$, are tiny). In the MF limit, one considers the initially disentangled state of $N_1$ identical particles (first component), and of $N_2$ identical particles (second component), with the initial wave function,

$$\psi(\vec{x}_1,...,\vec{x}_{N_1};\vec{y}_1,...\vec{y}_{N_2};0) = \psi_1(\vec{x}_1,0) \otimes ... \otimes \psi_1(\vec{x}_{N_1},0) \otimes \psi_2(\vec{y}_1,0) \otimes ... \otimes \psi_2(\vec{y}_{N_2},0).$$

(2.6)

It is shown in [8], that for any fixed dimensionless time, $\tau = \frac{\hbar}{m_1 r_0^2} t$, and $N_{1,2} \to \infty$, the wave function is,

$$\psi(\vec{x}_1,...,\vec{x}_{N_1};\vec{y}_1,...\vec{y}_{N_2};\tau) = \psi_1(\vec{x}_1,\tau) \otimes ... \otimes \psi_1(\vec{x}_{N_1},\tau) \otimes \psi_2(\vec{y}_1,\tau) \otimes ... \otimes \psi_2(\vec{y}_{N_2},\tau).$$

(2.7)

The single-particle wave functions, $\phi_1(\vec{x},\tau) = r_0^{3/2}\psi_1(\vec{x},\tau)$ and $\phi_2(\vec{y},\tau) = r_0^{3/2}\psi_2(\vec{y},\tau)$, for our Hamiltonian (2.3), satisfy the nonlinear coupled integro-differential equations of the Hartree-Fock type,

$$i\frac{\partial \phi_1(x,\tau)}{\partial \tau} = \left(-\frac{1}{2}\Delta_x - \int \frac{|\phi_1(x',\tau)|^2 d^3x'}{|x-x'|} - \beta \int \frac{|\phi_2(x',\tau)|^2 d^3x'}{|x-x'|}\right)\phi_1(x,\tau),$$

$$i\sqrt{\alpha}\frac{\partial \phi_2(x,\tau)}{\partial \tau} = \left(-\frac{1}{2}\Delta_x - \alpha\beta \int \frac{|\phi_2(x',\tau)|^2 d^3x'}{|x-x'|} - \alpha \int \frac{|\phi_1(x',\tau)|^2 d^3x'}{|x-x'|}\right)\phi_2(x,\tau), \quad (2.8)$$

$$\alpha = \left(\frac{m_2}{m_1}\right)^2, \quad \beta = \frac{M_2}{M_1}.$$

For simplicity, we omitted the vector symbol for coordinates in (2.8), and below. We also used a substitution in the second equation in (2.8): $y \to x$.

*Eigenvalue problem in MF limit.*

The corresponding eigenvalue problem, in the MF limit, can be formulated as follows. We assume that all axions, for both components, are in the same spherically symmetrical ground states which are described by the wave functions, $\phi_1(x)$ and $\phi_2(x)$. The wave functions satisfies the non-linear stationary equations of the Hartree-Fock type,

$$\varepsilon_1 \phi_1(x) = \left(-\frac{1}{2}\Delta_x - \int \frac{|\phi_1(x')|^2 d^3x'}{|x-x'|} - \beta \int \frac{|\phi_2(x')|^2 d^3x'}{|x-x'|}\right)\phi_1(x),$$

$$\varepsilon_2 \phi_2(x) = \left(-\frac{1}{2}\Delta_x - \alpha\beta \int \frac{|\phi_2(x')|^2 d^3x'}{|x-x'|} - \alpha \int \frac{|\phi_1(x')|^2 d^3x'}{|x-x'|}\right)\phi_2(x), \quad (2.9)$$

where $\varepsilon_{1,2}$ are the dimensionless eigenvalues. For convenience, we present here the relations between dimensional and dimensionless parameters which we use,

$$r_0 = \frac{\hbar^2}{Gm_1^2 M_1} - \text{characteristic size},$$

$$\vec{x} = \frac{\vec{r}}{r_0} - \text{dimensionless coordinate},$$

$$\tau = \frac{\hbar}{m_1 r_0^2}t - \text{dimensionless time}, \quad (2.10)$$

$$\phi_{1,2} = r_0^{3/2}\psi_{1,2} - \text{dimensionless wave functions},$$

$$\varepsilon_1 = \frac{m_1 r_0^2}{\hbar^2}E_1, \quad \varepsilon_2 = \frac{\sqrt{m_1 m_2}\, r_0^2}{\hbar^2}E_2 - \text{dimensionless energies}$$

In (2.10), $r_0$, is the gravitational MF analog of the Bohr radius in the hydrogen atom. The expressions in (2.9) can be simplified. Indeed, as it is well-known, the integration over polar and azimuthal angles in (2.9) can be performed explicitly. Using the expression,

$$\int_0^\pi d\theta \cdot \frac{\sin\theta}{\sqrt{x^2 - 2xx'\cos\theta + x'^2}} = \begin{cases} 2/x, & x > x', \\ 2/x', & x < x', \end{cases} \quad (2.11)$$

we obtain from (2.9) two coupled integro-differential equations for wave functions, $\phi_1(x)$ and $\phi_2(x)$, of the Hartree-Fock type,

$$-\frac{1}{2x}\frac{\partial^2}{\partial x^2}(x\phi_1(x)) - 4\pi\left(\frac{1}{x}\phi_1(x)\int_0^x s^2\phi_1^2(s)ds + \phi_1(x)\int_x^\infty s\phi_1^2(s)ds\right) -$$

$$4\pi\beta\left(\frac{1}{x}\phi_2(x)\int_0^x s^2\phi_2^2(s)ds + \phi_2(x)\int_x^\infty s\phi_2^2(s)ds\right) = \varepsilon_1\phi_1(x),$$

$$-\frac{1}{2x}\frac{\partial^2}{\partial x^2}(x\phi_2(x)) - 4\pi\alpha\beta\left(\frac{1}{x}\phi_2(x)\int_0^x s^2\phi_2^2(s)ds + \phi_2(x)\int_x^\infty s\phi_2^2(s)ds\right) - \quad (2.12)$$

$$4\pi\alpha\left(\frac{1}{x}\phi_1(x)\int_0^x s^2\phi_1^2(s)ds + \phi_1(x)\int_x^\infty s\phi_1^2(s)ds\right) = \varepsilon_2\phi_2(x).$$

Also, we add two standard conditions,

$$\phi_{1,2}(x=\infty) = 0,$$

$$4\pi\int_0^\infty x^2\phi_{1,2}^2(x)dx = 1. \quad (2.13)$$

Using the substitutions,

$$\varphi_{1,2}(x) = \sqrt{4\pi}\,x\phi_{1,2}(x), \quad (2.14)$$

we obtain from (2.12) the equations which include the effective potentials,

$$\varepsilon_1 \varphi_1(x) = -\frac{1}{2}\frac{\partial^2 \varphi_1(x)}{\partial x^2} + U_1(x)\varphi_1(x),$$

$$U_1(x) = -\left(\frac{1}{x}\int_0^x \varphi_1^2(s)ds + \int_x^\infty \frac{\varphi_1^2(s)}{s}ds\right) - \beta\left(\frac{1}{x}\int_0^x \varphi_2^2(s)ds + \int_x^\infty \frac{\varphi_2^2(s)}{s}ds\right),$$

$$\varepsilon_1 = \frac{1}{2}\int_0^\infty \left(\frac{\partial \varphi_1(x)}{\partial x}\right)^2 dx + \int_0^\infty U_1(x)\varphi_1^2(x)dx,$$

$$\varepsilon_2 \varphi_2(x) = -\frac{1}{2}\frac{\partial^2 \varphi_2(x)}{\partial x^2} + U_2(x)\varphi_2(x),$$

$$U_2(x) = -\alpha\left(\frac{1}{x}\int_0^x \varphi_1^2(s)ds + \int_x^\infty \frac{\varphi_1^2(s)}{s}ds\right) - \alpha\beta\left(\frac{1}{x}\int_0^x \varphi_2^2(s)ds + \int_x^\infty \frac{\varphi_2^2(s)}{s}ds\right),$$

$$\varepsilon_2 = \frac{1}{2}\int_0^\infty \left(\frac{\partial \varphi_2(x)}{\partial x}\right)^2 dx + \int_0^\infty U_2(x)\varphi_2^2(x)dx,$$

$$\varphi_{1,2}(0) = \varphi_{1,2}(\infty) = 0,$$

$$\int_0^\infty \varphi_{1,2}^2(x)dx = 1. \tag{2.15}$$

### III. Numerical Protocol

The ground state solutions of the system of Eqs. (2.9) were found numerically as stationary solutions of the system of effective dynamic equations,

$$i\frac{\partial \phi_1(x,t)}{\partial t} = \left(\frac{1}{2}\Delta_x + V_1(x,t) + \beta V_2(x,t) + \varepsilon_1\right)\phi_1(x,t) \equiv \hat{H}_1[V_1, \beta V_2; \varepsilon_1]\phi_1(x,t), \tag{3.1}$$

$$i\frac{\partial \phi_2(x,t)}{\partial t} = \left(\frac{1}{2}\Delta_x + \alpha V_1(x,t) + \alpha\beta V_2(x,t) + \varepsilon_2\right)\phi_2(x,t) \equiv \hat{H}_2[\alpha V_1, \alpha\beta V_2; \varepsilon_2]\phi_2(x,t). \tag{3.2}$$

$$V_1(x,t) = \int_0^\infty \frac{|\phi_1(x',t)|^2 d^3x'}{|x-x'|}, \quad V_2(x,t) = \int_0^\infty \frac{|\phi_2(x',t)|^2 d^3x'}{|x-x'|}. \tag{3.3}$$

The implicit finite-difference method of the first-order in time was used with the discretization steps in time and space, $\Delta t$ and $\Delta x$, correspondingly. This method is not conservative for parabolic type of equations, (3.1), (3.2). However, it is precisely its dissipative properties, described in detail in our work [7], that ensure the transition of the considered system of axions from a certain initial state to the desired stationary state. The time of this transition and the corresponding stationary solution depend on the initial condition. It should be chosen carefully to avoid divergences in the described scheme, as well as solutions corresponding to the excited states of the system. The method, which we have chosen that addresses these problems, is presented below.

Suppose that the stationary solution of the system, $\phi_{1,0}(x)$, $\phi_{2,0}(x)$, $\varepsilon_{1,0}$, $\varepsilon_{2,0}$, for parameters, $\alpha_0$ and $\beta_0$ is known. It is necessary to find a solution for: $\alpha_1 = \alpha_0 + \Delta\alpha$, $\beta_1 = \beta_0 + \Delta\beta$. The sequence of operations, used in the iterative algorithm, is as follows (the second index of the functions and values, presented above and below, indicates the number of performed iterations, $k = 1, 2, 3, ..., K$ ):

1. Find the solution of Eq. (3.1),

$$i \frac{\partial \phi_{1,k}(x,t)}{\partial t} \equiv \hat{H}_1 \left[ V_{1,k-1}, \beta_1 V_{2,k-1}; \varepsilon_{1,k-1} \right] \phi_{1,k}(x,t), \text{ after } n\text{-time steps.}$$

2. Normalize, $\phi_{1,k}(x,t)$, and calculate/redefine, $V_{1,k}, \varepsilon_{1,k}$.

2. Find the solution of Eq. (3.2),

$$i \frac{\partial \phi_{2,k}(x,t)}{\partial t} \equiv \hat{H}_2 \left[ \alpha_1 V_{1,k}, \alpha_1 \beta_1 V_{2,k-1}; \varepsilon_{2,k-1} \right] \phi_{2,k}(x,t), \text{ after } n\text{-time steps.}$$

4. Normalize the obtained function, $\phi_{2,k}(x,t)$, calculate, $V_{2,k}$ and $\varepsilon_{2,k}$, for the following, $(k+1)$-th iteration, and repeat the sequence of steps 1-4.

In our simulations, we have chosen the domain in $x$: $0 \leq x \leq X = 15$ (with boundary conditions, $\phi_1(x = X) = \phi_2(x = X) = 0$), $\Delta x = X / 5 \times 10^4$, $\Delta t = 0.8$, $n = 200$, $K = 15$, $\Delta \alpha = 0.2$, $\Delta \beta = 0.002$. With such discreteness of parameters, $\alpha$ and $\beta$, the calculations of stationary states of the system along straight lines, $1 \leq \alpha \leq 100$ ($\beta = const$) or $1 \leq \beta \leq 1$ ($\alpha = const$) takes 2 hours of operation of the standard 3.2 GHz-desktop. As a result, for each set of parameters, $\alpha$ and $\beta$, the energy values, $\varepsilon_1$ and $\varepsilon_2$, are computed with a relative error, $\Delta \varepsilon_{1,2} / \varepsilon_{1,2} < 10^{-6}$. As the "starting state" for calculating an arbitrary "trajectory" of the system on the plane, $(\alpha, \beta)$, it is convenient to choose the solution of the system for $\alpha = 1$. In this case, the functions, $\phi_1(x) = \phi_2(x)$ can be easily found from the equation,

$$i \frac{\partial \phi_1(x,t)}{\partial t} = \left( \frac{1}{2} \Delta_x + (1+\beta) V_1(x,t) + \varepsilon_1 \right) \phi_1(x,t), \tag{3.4}$$

whose stationary solution, with physically justified initial approximations, is easily found by the iteration method described above (for details see ref. [7]).

### IV. Results of Numerical Simulations

In this section, we present the results of our numerical simulations. We compute the radial distribution of the mass density of LA and design the parameters which can be used in order to distinguish a two-component DMH from a one-component DMH.

As will be discussed below, under all values of parameters, $\beta = M_2/M_1$ and $\alpha = (m_2/m_1)^2$, the region of localization of the heavy axions (HA) is narrower than the region of localization of the light axions (LA). Depending on the ratio of total masses of axionic sub-systems, $\beta$, two limiting cases of the HA concentration, at the center of the system, can be realized. In the first case, the factor of "compression" of the HA is mainly caused by the potential produced by the LA, the spatial distribution of which is weakly disturbed by the presence of a HA. In the second case, the spatial distribution of HA is determined mainly by its own potential, which also significantly narrows the region of localization of LA. It is evident that the first case can be realized only at small $\beta \ll 1$. In Fig. 1, we present three probability density functions, $p_1(x) = 0.5(1+q)|\phi_1(x)|^2$, $p_2(x) = 0.5(1-q)|\phi_2(x)|^2$, and $P(x)$,

$$P(x) = (M_1/M)p_1(x) + (M_2/M)p_2(x) = p_1(x) + p_2(x),$$
$$p_{1,2}(x) = |\phi_{1,2}(x)|^2, \quad p_1(x) = 0.5 \times (1+q)p_1(x), \quad p_2(x) = 0.5 \times (1-q)p_1(x), \tag{4.1}$$
$$M_1 + M_2 = M, \quad \frac{M_2}{M_1} = \frac{1-q}{1+q} = \beta.$$

In (4.1), the parameter, $q$, is introduced for convenience.

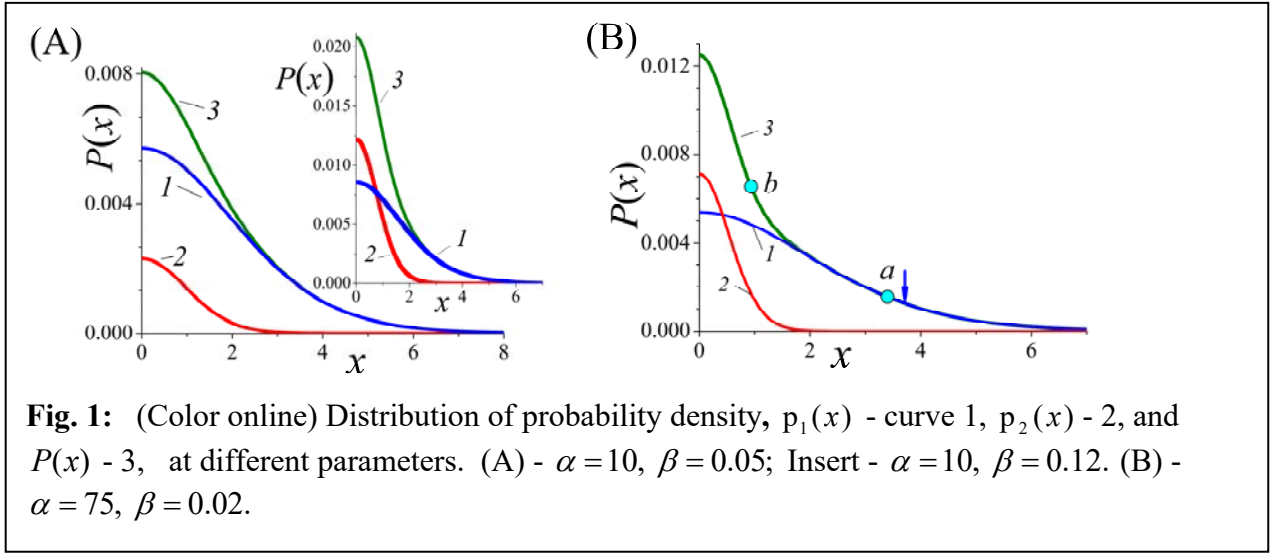

**Fig. 1:** (Color online) Distribution of probability density, $p_1(x)$ - curve 1, $p_2(x)$ - 2, and $P(x)$ - 3, at different parameters. (A) - $\alpha = 10$, $\beta = 0.05$; Insert - $\alpha = 10$, $\beta = 0.12$. (B) - $\alpha = 75$, $\beta = 0.02$.

Depending on the behavior of $P(x)$, one can detect the presence of the second component of DMH. For example, the formation of the distribution, $P(x)$, from two different-scale distributions, $p_1(x)$ and $p_2(x)$ (Fig. 1B), is clearly manifested in the existence of two points (*a* and *b*), where the absolute value of the negative curvature of $P(x)$ is at maximum. The arrow indicates the extremum point of negative curvature of $p_1(x)$ in the case $\beta = 0$. Below, we design the integral criteria for detection of the heavy component.

*Integral criteria for detecting of HA.*

For future analysis, we present here the main parameters of a one-component system of LA: $\alpha = 1$ and the total "dimensionless mass", $(M_1 + M_2)/M_1 = (1+\beta)$,

$$\varepsilon_{1,0} = -0.16277 \times (1+\beta)^2, \quad V_{1,0} = -0.3155 \times (1+\beta),$$
$$|\phi_{1,0}(x=0)|^2 = 0.0044 \times (1+\beta)^3, \quad \hat{x}_{v1,0} = 7/(1+\beta). \tag{4.2}$$

Here, $\varepsilon_{1,0}$ - is the eigenenergy of the system, $V_{1,0}$ - is the depth of the potential well, $\hat{x}_{v1,0}$ - is the characteristic width of the spatial distribution of the axion density, which is determined from the condition: $|\phi_1(x=\hat{x}_{v1,0})|^2 \approx 10^{-2} \times |\phi_1(x=0)|^2$. The shape of this distribution (see curve 1 in Fig. 3) can be well represented by the function,

$$|\phi_1(x)|^2 \approx |\phi_{1,0}(x=0)|^2 \times \exp(-x^2(1+\beta)^2/11). \tag{4.3}$$

The potential (3.3), created by such distribution, is well approximated by the function,

$$V_1(x) = V_{1,0}/[1+x^2(1+\beta)^2/39], \quad 0 \leq x \leq \hat{x}_{v1,0},$$
$$V_1(x) = -1/x, \quad x > \hat{x}_{v1,0}.$$

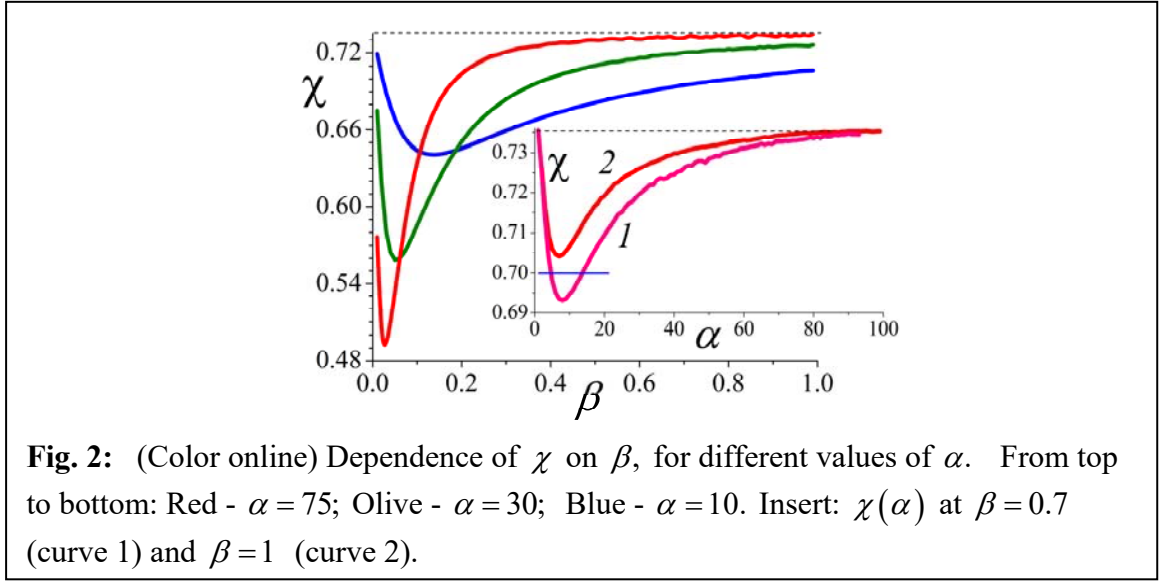

**Fig. 2:** (Color online) Dependence of $\chi$ on $\beta$, for different values of $\alpha$. From top to bottom: Red - $\alpha = 75$; Olive - $\alpha = 30$; Blue - $\alpha = 10$. Insert: $\chi(\alpha)$ at $\beta = 0.7$ (curve 1) and $\beta = 1$ (curve 2).

The point, $x = \hat{x}_{v1,0}$, defines the boundary beyond which the axionic potential is equal, approximately, to the potential of a point mass.

If the function, $P(x)$, is determined from observations, then the existence of the second component can be established on the basis of knowledge of physical factors that determine the shape of this function. Let us illustrate the problems on the example of constructing the simplest of all possible indicators, $\chi$, of two-component system.

To calculate, $\chi$, define the value, $x_w$, from the relation, $P(x_w) = P(x=0)/2$. Then,

$$\chi = \int_0^{x_w} P(x)dx \bigg/ \int_0^{\infty} P(x)dx = 1/(1 + I_b/I_a). \tag{4.4}$$

$$I_a = \int_0^{x_w} P(x)dx, \quad I_b = \int_{x_w}^{\infty} P(x)dx. \tag{4.5}$$

For a single-component system $(\alpha = 1)$, we have: $\chi_0 = 0.735$, $I_b/I_a \approx 0.36$, independently of the value of $\beta$. Consider in detail the physical factors determining the evolution of the dependence $\chi(\beta)$, for a given parameter $\alpha > 1$.

If the total mass, $M_2$, of the HA is small $(\beta \ll 1)$, then the potential energy of HA, $u_{p,2}$, is determined mainly by the potintial of LA, $V_1(\vec{x})$,

$$u_{p,2} \approx -\alpha V_1(\vec{x}), \quad V_1(\vec{x}) = \int_0^x |\phi_1(\vec{x}')|^2 d^3\vec{x}'/|\vec{x} - \vec{x}'|, \tag{4.6}$$

which react poorly on the heavy component due to its low total mass, $M_2$. In Fig. 1, curves 1 correspond rather well to the parameters of the single-component distribution. However, HA with less potential energy, $u_{p,2} \approx -\alpha V_1(\vec{x}) < -V_1(\vec{x}) = u_{p,1}$, are concentrated at the center of the system (curves 2 in Fig. 1) and cause small increase in the density, $|\phi_1(x \approx 0)|^2 > 0.0044$, of LA.

The distribution, $P(x)$, represented by curve 3 in Fig. 1B, cannot be satisfactorily approximated by a function of the form (4.3). This is caused by the pronounced downward deflection of this function. It is this deflection that indicates the two-component DMH. The anomaly of such a deflection is more difficult to visually notice for curves 3 in Fig. 1A. However, this can be identified by comparing the value of the multi-component indicator, $\chi$, with the value, $\chi_0 = 0.735$, determined for a single-component system (difference $\Delta_\chi = \chi_0 - \chi > 0$, - see Fig. 2).

The principle of detecting HA is easy to follow according to Fig. 3. The increase in the curve deflection, $\hat{P}(x) = P(x)/P(x=0)$, with growth of $\beta$ (mass of HA) reduces the distribution width, $x_w$. The ratio, $I_b/I_a$, increases ($\chi$ decreases), although each of the integrals, $I_a$ and $I_b$, decreases relative to the corresponding values obtained for the "reference" dependence — curve 1 in Fig. 3.

As one can see from the results of Fig. 2, the indicator, $\chi$, for the distributions, $P(x)$, shown in Fig. 1, is close to the minimum values, $\chi(\beta)$, for $\alpha = 10$ and 75. We also note an increase of $\Delta_\chi = \chi_0 - \chi$, with an increase in the relative mass of HA, $\alpha$, at a fixed value of their total mass, $\beta \ll 1$ (see Fig. 2). The physical factor, $\chi(\alpha)\big|_{\beta \ll 1}$, responsible for such changes, can be easily imagined qualitatively. The larger the mass of an individual particle, the narrower the region of their localization in the field of LA, and the higher the local increase of $P(x \approx 0)$ with an unperturbed peripheral zone. So, a detection of the curve deflection becomes more reliable. A similar effect is observed with an increase in the total mass of HA, $\beta$, if it is small enough (see in Fig. 2 the regions where $\partial \chi / \partial \beta < 0$).

However, with further growth of $\beta$, the intrinsic field of HA becomes dominant, and the second of the above limiting options for the formation of the state of the system is realized. The distribution density of axions, $P(x)$, in the central region, is determined mainly by a heavy component, the shape of which is difficult to distinguish from a one-component distribution (2) - see the inset in Fig. 3. The width, $\hat{x}_v$, of the presented distribution, can be fairly accurately estimated by neglecting the field of the LA. From the corresponding relation in Eq. (4.3), it follows that $\hat{x}_v \approx 7/(\alpha\beta) = 7/(75 \times 0.6) \approx 15.6$ (in relations (4.3), instead of a quantity $(1+\beta)$, the product, $\alpha\beta$, should be used). So, the visual radius is reduced by ~ 45 times compared with the "reference" value (in good agreement with the numerically obtained result, see Fig. 3). The energy of HA, $\varepsilon_2 = -397.77$, and its estimate, according to the scaling rules presented in Eq. (4.1), is:

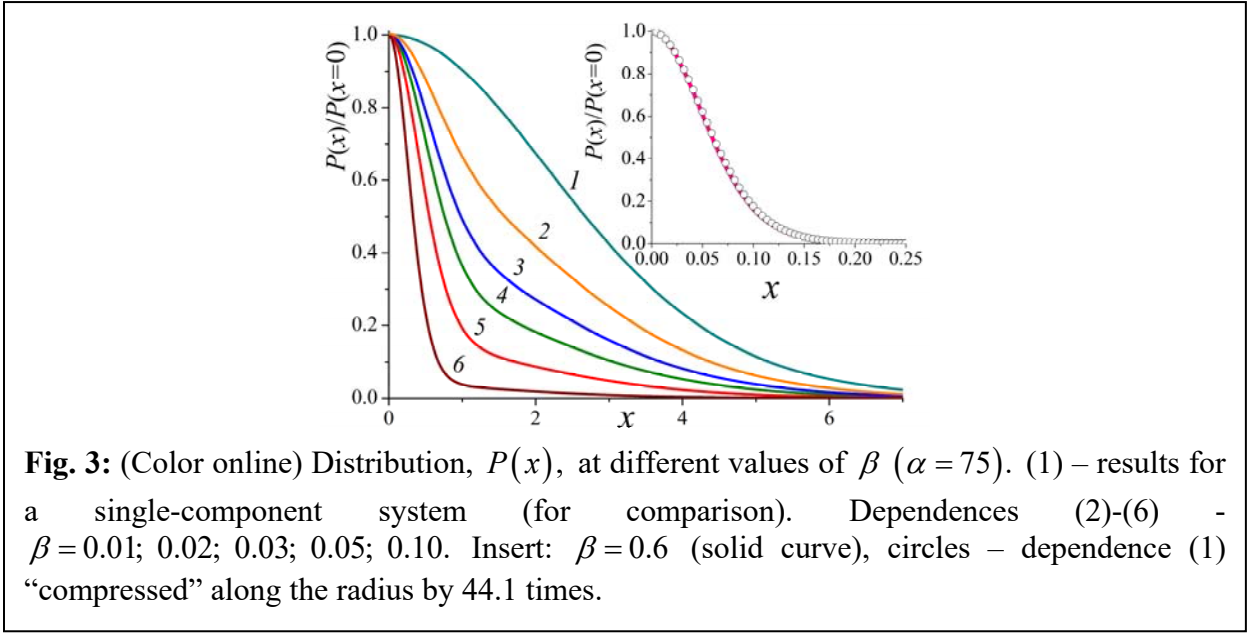

**Fig. 3:** (Color online) Distribution, $P(x)$, at different values of $\beta$ ($\alpha = 75$). (1) – results for a single-component system (for comparison). Dependences (2)-(6) - $\beta = 0.01; 0.02; 0.03; 0.05; 0.10$. Insert: $\beta = 0.6$ (solid curve), circles – dependence (1) "compressed" along the radius by 44.1 times.

$$\tilde{\varepsilon}_2 = \varepsilon_{1,0} \times (\alpha\beta)^2 \approx -330.$$

The mentioned similarity of the analyzed distribution with the one-component distribution leads to a vanishingly small value, $\Delta_\chi = \chi_0 - \chi$ - see Fig. 2. Despite the large total mass, $M_1$, of LA ($M_1/M_2 = 1/\beta \approx 1.67$), they became non-detectable. In the experiment, reliable recording of $\Delta_\chi$ is limited to a certain minimum value. If we assume that $\Delta_{\chi,\min} = 0.05\chi_0$, then the condition $\chi < \chi_0 - \Delta_{\chi,\min} \approx 0.7$, is satisfied in a narrow range of $\beta$, and the width of this range narrows sharply with increasing of $\alpha$. For example, at $\alpha = 75$, registration of HA by the described method is possible only in the range, $0.01 < \beta < 0.18$  $0.01 < \beta < 0.18$ – see Fig. 2. If the total masses of LA and HA are equal $(\beta=1)$, then the detection of such a state is impossible at any values of $\alpha>1$ - see the inset in Fig. 2, curve 2.

Although at smaller value of $\beta$ this detection is possible (curve 1). The origins of the problem are discussed in the next sub-section. They are due to the fact that the spatial distributions of LA and HA are sharply separated with certain combinations of system parameters, and the mutual influence between them is significantly weakened.

*Effects of the constriction of heavy axions.*

The characteristic of the spatial localization of the DMH components is the coordinates of their

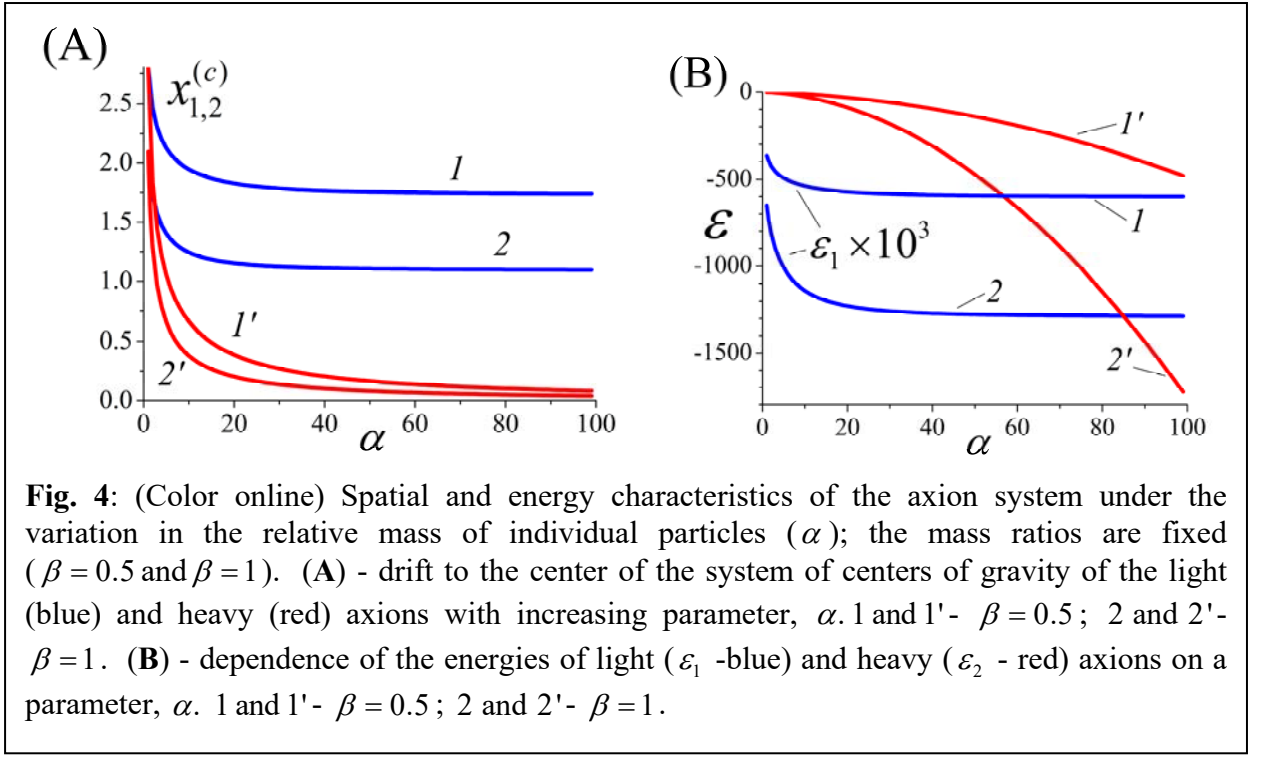

**Fig. 4**: (Color online) Spatial and energy characteristics of the axion system under the variation in the relative mass of individual particles ($\alpha$); the mass ratios are fixed ($\beta=0.5$ and $\beta=1$). (**A**) - drift to the center of the system of centers of gravity of the light (blue) and heavy (red) axions with increasing parameter, $\alpha$. 1 and 1'- $\beta=0.5$; 2 and 2'- $\beta=1$. (**B**) - dependence of the energies of light ($\varepsilon_1$ -blue) and heavy ($\varepsilon_2$ - red) axions on a parameter, $\alpha$. 1 and 1'- $\beta=0.5$; 2 and 2'- $\beta=1$.

"centers of gravity",

$$x_{1,2}^{(c)} = \int_0^\infty |\phi_{1,2}(x)|^2 x\,dx. \qquad (4.7)$$

Fig. 4A, represents the dependences of these values on $\alpha$, at fixed values of $\beta \geq 0.5$. It is seen that the "overlap" of the densities of light, $|\phi_1(x)|^2$ and heavy $|\phi_2(x)|^2$, axions ($x_1^{(c)}$ of order $x_2^{(c)}$) is realized only at relatively small values of $\alpha$ ($\alpha<10$). The initial increase in the mass of HA is accompanied by both their concentration at the center of the system and the entrainment to the center of LA. However, for $\alpha > 10-15$, a compression of the light component practically ceases when the constriction of the heavy component continues. The ratio, $\eta(\alpha) = x_1^{(c)}/x_2^{(c)}$, rises sharply to 20-25 almost linearly in $\alpha$,

$$\eta(\alpha) = 1 + \hat{\eta} \times (\alpha - 1), \tag{4.8}$$

where $\hat{\eta} = 0.20$ and $\hat{\eta} = 0.255$ for $\beta = 0.5$ and 1, correspondingly. In this case, the energy levels of the light component are practically stabilized (see Fig. 4B), and the potential, $U_1(x)$, for light component,

$$U_1(\vec{x}) = V_1(\vec{x}) + \beta V_2(\vec{x}), \quad V_2(\vec{x}) = \int_0^x |\phi_2(\vec{x}')|^2 d^3\vec{x}' / |\vec{x} - \vec{x}'|, \tag{4.9}$$

does not change. The physical mechanism of stabilization of $x_1^{(c)}$ and $\varepsilon_1$ is easy to interpret. The concentration of HA increases with growth of $\alpha$ within a sphere of small radius, $x_{v2}$. This leads to the fact that the bulk of LA remains in a vast zone outside this sphere, $x > x_{v2}$. For them, the potential created by HA does not change with a further decrease of $x_{v2}$: $V_2(x) = -1/x$, $x > x_{v2}$. That is, for the sub-system of LA, an almost stationary state is established. Small changes in its parameters are associated only with changes directly in the narrow zone of localization of HA.
With strong constriction of HA, not only their effect on LA is weakened (in the above sense). The parameters of the HA themselves can be fairly accurately estimated if we neglect the field of LA and consider a one-component state with potential, $U_2(\vec{x}) \approx -\alpha\beta V_1(\vec{x})$. Then (see the relations (4.1)), the estimate, $\varepsilon_2(\alpha) \approx -0.163(\alpha\beta)^2$, is in good agreement with the exact solutions (see the curves 1' and 2' in Fig. 4B) for large $\alpha$.

An idea of the evolution of the characteristics of the system with an increase in the relative mass, $\beta$, of HA ($\alpha$ is fixed) is given in Fig. 5.

A small difference between the curves (3) and (4) in Fig. 5A and Fig. 5C reflects that in Fig. 4 values, $x_1^{(c)}(\alpha)$ and $\varepsilon_1(\alpha)$, practically do not change for $\alpha > 20$ as a result of the pronounced spatial separation of LA and HA. Under such conditions, the two-component parameter, $\chi$, used above, "fades", and does not distinguish between the divisions that has arisen.

On the other hand, the data of Fig. 2 are very informative. The family of "resonant" dependencies, $\chi(\beta)$ and $\chi(\alpha)$, indicates that a significant spatial separation of axion subsystems develops to the right of the region of resonance. It occurs approximately in the zone: $\alpha > 20$ and $\beta > 0.2$. A good accuracy of such an estimate can be seen by analyzing the data of Fig. 4 and Fig. 5.

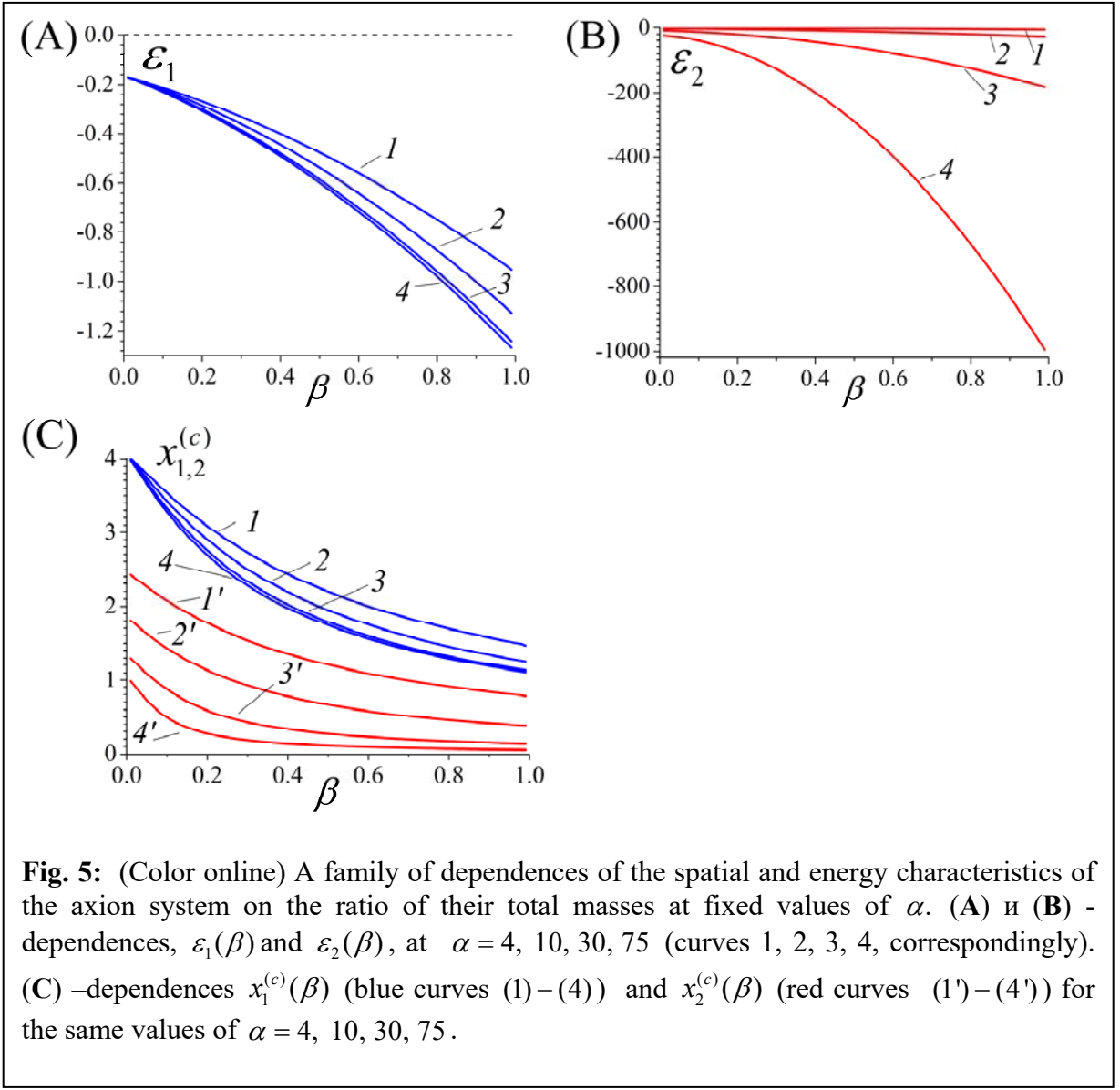

**Fig. 5:** (Color online) A family of dependences of the spatial and energy characteristics of the axion system on the ratio of their total masses at fixed values of $\alpha$. **(A)** и **(B)** - dependences, $\varepsilon_1(\beta)$ and $\varepsilon_2(\beta)$, at $\alpha = 4, 10, 30, 75$ (curves 1, 2, 3, 4, correspondingly). **(C)** –dependences $x_1^{(c)}(\beta)$ (blue curves (1)–(4)) and $x_2^{(c)}(\beta)$ (red curves (1')–(4')) for the same values of $\alpha = 4, 10, 30, 75$.

The physical principles for constructing a more accurate multicomponent index, $\hat{\chi}$, are shown in Fig. 6. In Fig. 6A, the stratification of axion subsystems still allows the possibility of detecting two-components by means of parameter $\chi$ ($\chi \approx 0.69$). The density of axions at the point of maximum negative curvature of the function, $P(x) = 0.5(1+q)\left[|\phi_1(x)|^2 + \beta|\phi_2(x)|^2\right]$, is of the same order of magnitude as $P(x=0)$. This indicates still a significant overlap of the distribution, $|\phi_1(x)|^2$, with the distribution, $|\phi_2(x)|^2$ (usually the point of maximum curvature, $x = x_{\max}^{(curv)}$, lies in the zone where $|\phi_1(x)|^2 \approx |\phi_2(x)|^2$). In Fig. 6B, the point, $x = x_{\max}^{(curv)}$, is pushed into the region of vanishingly low density of axions, i.e. overlapping distributions of axion subsystems is minimal (Fig. 6C).

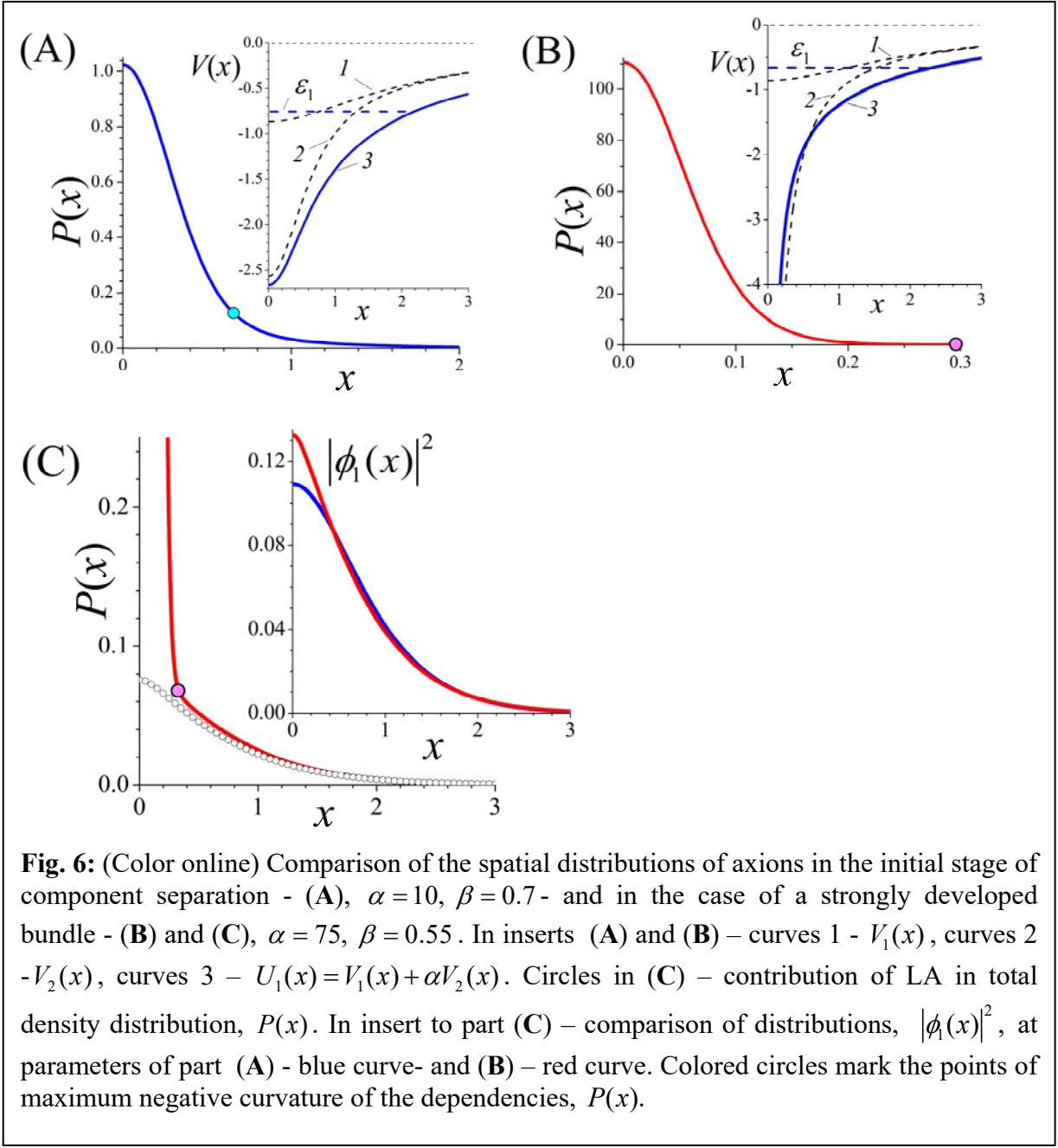

**Fig. 6:** (Color online) Comparison of the spatial distributions of axions in the initial stage of component separation - **(A)**, $\alpha = 10$, $\beta = 0.7$ - and in the case of a strongly developed bundle - **(B)** and **(C)**, $\alpha = 75$, $\beta = 0.55$. In inserts **(A)** and **(B)** – curves 1 - $V_1(x)$, curves 2 - $V_2(x)$, curves 3 – $U_1(x) = V_1(x) + \alpha V_2(x)$. Circles in **(C)** – contribution of LA in total density distribution, $P(x)$. In insert to part **(C)** – comparison of distributions, $|\phi_1(x)|^2$, at parameters of part **(A)** - blue curve- and **(B)** – red curve. Colored circles mark the points of maximum negative curvature of the dependencies, $P(x)$.

This statement could be predicted based on estimates. The radius of the sphere of localization of HA in Fig. 6A, $x_{v2} \approx 7/(\alpha\beta) = 1$ and in Fig. 6B - $x_{v2} \approx 7/(\alpha\beta) = 0.17$, while the distribution,

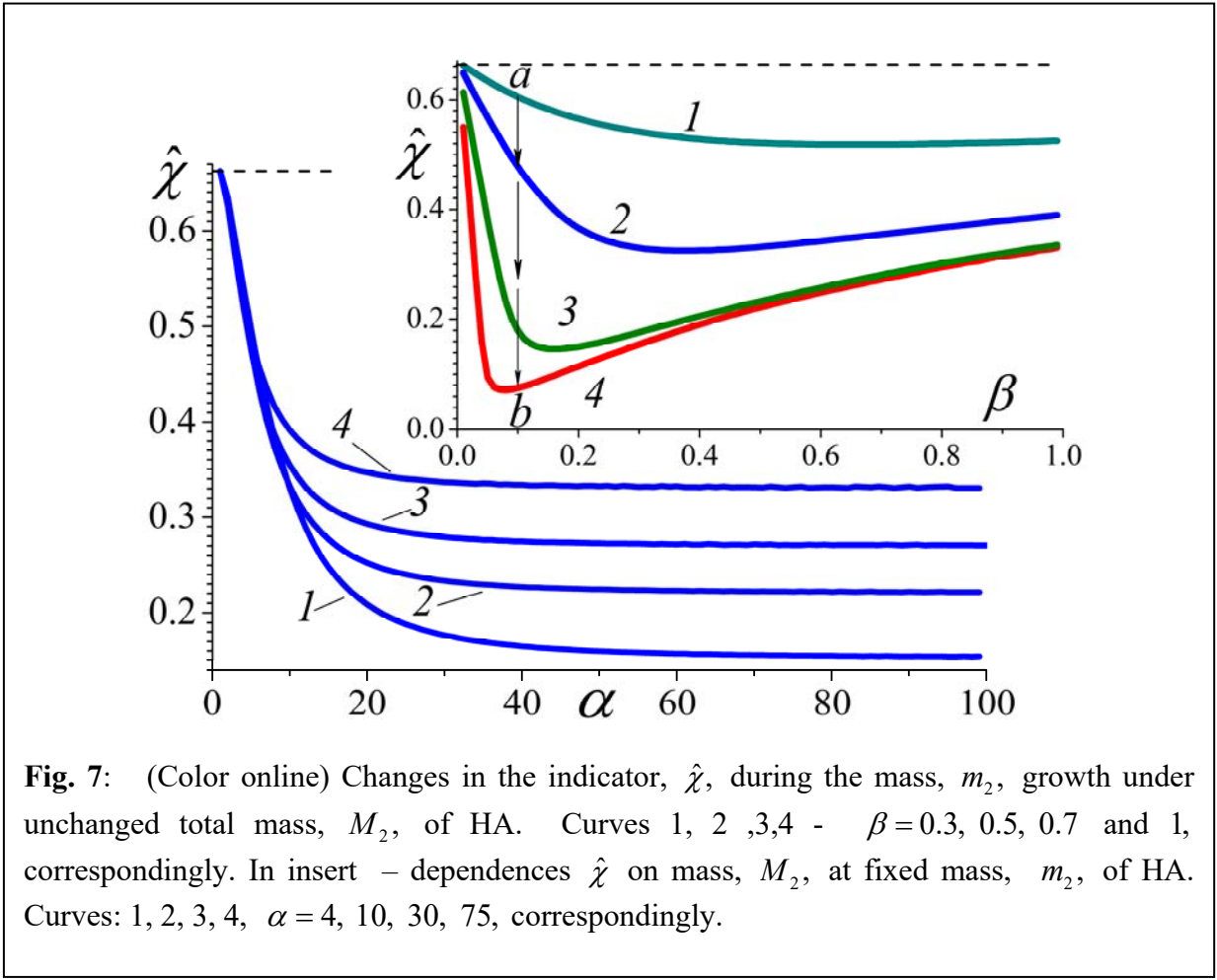

**Fig. 7:** (Color online) Changes in the indicator, $\hat{\chi}$, during the mass, $m_2$, growth under unchanged total mass, $M_2$, of HA. Curves 1, 2, 3, 4 - $\beta = 0.3, 0.5, 0.7$ and 1, correspondingly. In insert – dependences $\hat{\chi}$ on mass, $M_2$, at fixed mass, $m_2$, of HA. Curves: 1, 2, 3, 4, $\alpha = 4, 10, 30, 75$, correspondingly.

$|\phi_1(x)|^2$, remains virtually unchanged (see the insets in Figs. 6A, B, C).

Thus, the problem of recognizing of the two-component system of axions reduces to the problem of detecting the halo from LA around a dense nucleus from HA, the density of which is many times (several orders of magnitude - see Figs. 6B, C) higher than the density of the light component. In this case, the mass ratio of individual particles can be within 10 ($\alpha < 100$). If the registration of densities differing by orders of magnitude is possible, then the indicator, $\hat{\chi}$, can be constructed according to the principle of Eqs. (4.4) and (4.5), in which the integrals,

$$I_a = \int_0^{x_w} x^2 P(x)dx, \quad I_b = \int_{x_w}^{\infty} x^2 P(x)dx, \quad P(x_w) = P(x=0)/e^2, \tag{4.10}$$

determine the masses of axions in the sphere of radius, $x_w$, and outside it, $x > x_w$. For a one-component system, $I_b / I_a \approx 0.51$, and the reference value of the indicator is: $\hat{\chi}_0 = 1/(1 + I_b / I_a) \approx 0.662$.

In the case of pronounced separation of axions in a two-component system, $\hat{\chi} < \hat{\chi}_0$, due to an increase in the ratio, $I_b / I_a$. The first integral represents the bulk of the halo mass, i.e. the bulk of the mass of LA. For example, in Fig. 6B, $x_w \approx 0.1$, and the integral, $I_b \left( x \geq x_w \approx 0.1 \right)$ almost completely includes LA (see the curve presented by circles in Fig. 6C) and a part of the core. The second integral to a lesser extent represents the mass of the nucleus of HA, which is initially less than the mass of LA. As a result, the expected value of $I_b / I_a > 1$ and $\hat{\chi} < \hat{\chi}_0$. The construction principle of $\hat{\chi}$ itself leads to the conclusion that, for a fixed parameter, $\beta$, the quantity, $\hat{\chi}(\alpha)$, reaches saturation with growth of $\alpha$.
This is confirmed by the results presented in Fig. 7. (Compare the data in this figure with the insert data in Fig. 2.) Note that in the saturation zone $\hat{\chi}(\alpha; \beta = 0.7) < \hat{\chi}(\alpha; \beta = 1)$. This relation is associated with a decrease of $I_a$ relatively to $I_b$, since the mass of the halo is greater than the mass of the nucleus.

Dependencies, shown in the inset to Fig. 7, demonstrate a high performance indicator, $\hat{\chi}$. In the region, $\beta \ll 1$, the results of this indicator correspond to the results of Fig. 2, however, at large $\beta$ it is noticeably superior in sensitivity to the two-component system. Moreover, the alternation of blue-olive-red curves in the region, $\beta > 0.2$, occurs in the reverse order than in Fig. 2. We also note a significant drop of $\hat{\chi}$ along the straight line $ab$ (the inset in Fig. 7, $\beta = 0.1$), which has a simple and clear interpretation. HA with a small total mass, $M_2$, slightly perturb the spatial distribution of the light component. But they themselves compress with increasing $m_2$ ($\alpha \gg 1$) into a sphere of small radius with a high concentration (see, for example, curve 6 in Fig. 3). In this case, the ratio, $I_b / I_a \Rightarrow M_1 / M_2 = 1/\beta = 10$. Accordingly, the estimate of the minimum value of, $\hat{\chi} \approx 1/11 \approx 0.9$, is close to the exact result obtained.

**V. Relations to Hawking Temperature and Entropy and to Unruh Temperature**

For the purposes of illustration, consider here only a MF solution with one-component ULA in BEC state [7]. (See Fig. 8.)

*Relation to the Hawking temperature.*

We have for a single-particle dimensional energy of the BEC ground state [7],

$$E = -\varepsilon \frac{\hbar^2}{m r_0^2}, \tag{5.1}$$

where, according to [7], $\varepsilon = 0.16277$. In (5.1), $E$ and $r_0$, are the gravitational MF analogs of the ground state energy and the Bohr radius in the hydrogen atom, correspondingly.

In the MF limit, used here, the only characteristic dimensional spatial scale is $r_0$, defined in (2.10). On the other hand, a gravitational system has another characteristic dimensional spatial parameter,

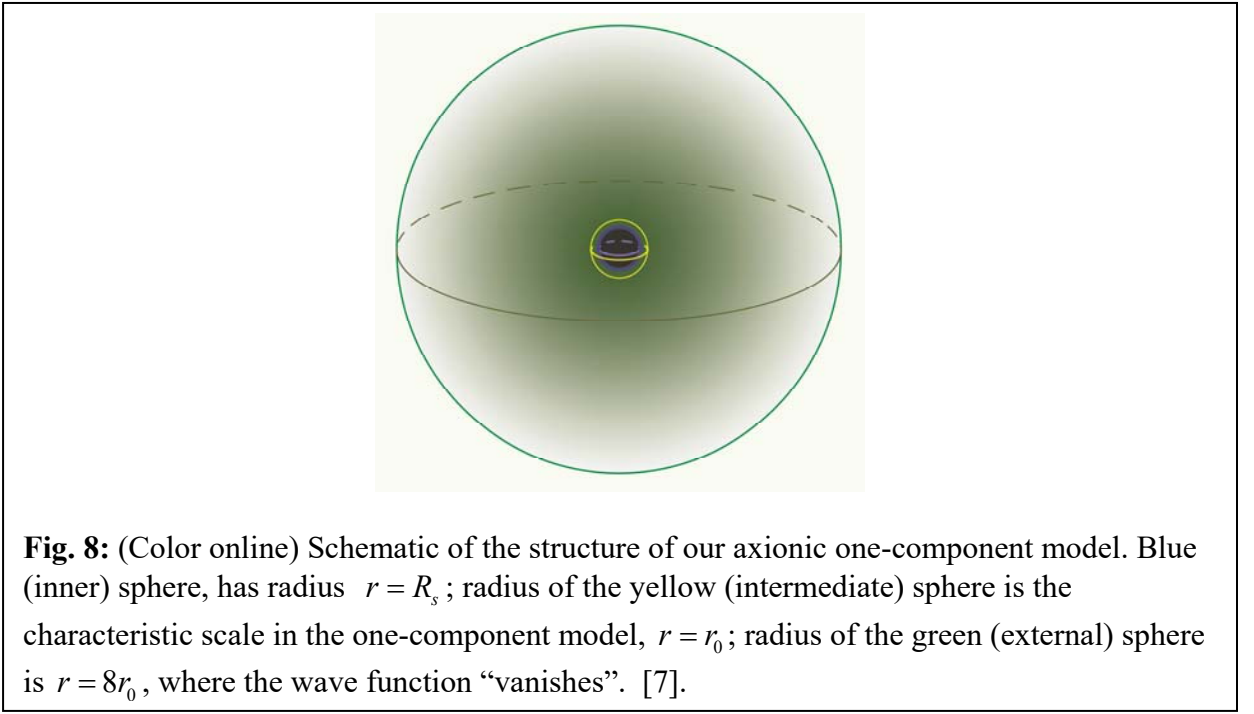

**Fig. 8:** (Color online) Schematic of the structure of our axionic one-component model. Blue (inner) sphere, has radius $r = R_s$; radius of the yellow (intermediate) sphere is the characteristic scale in the one-component model, $r = r_0$; radius of the green (external) sphere is $r = 8r_0$, where the wave function "vanishes". [7].

the Schwarzschild radius,

$$R_s = \frac{2GM}{c^2}. \tag{5.2}$$

We use the following relation between $r_0$ and $R_s$,

$$\chi_{BH} r_0 = R_s; \quad (\chi_{BH} < 1), \tag{5.3}$$

where, $\chi_{BH}$, is the dimensionless factor. The condition, $\chi_{BH} < 1$, is required because in the MF limit, used here, the size of the corresponding BH $(R_s)$ is less than the characteristic size of the MF wave function, $r_0$. From (2.10) and (5.3), we have the relation between $m$ and $M$,

$$m = \sqrt{\frac{\chi_{BH}}{2}} \frac{c\hbar}{GM}. \tag{5.4}$$

If follows from (5.4), that for a given axion mass, $m$, the dimensionless parameter, $\chi_{BH}$, depends only on the total mass, $M$. Using (5.4), we can exclude $m$ in (5.1). We have,

$$E = -\pi\varepsilon(2\chi_{BH})^{3/2} \cdot k_B T_H; \quad \left(k_B T_H = \frac{\hbar c^3}{8\pi GM}\right), \tag{5.5}$$

where $T_H$ is the Black Hole Hawking temperature [9,10]. So, the ground state energy, $|E|$ (5.1), differs from the Hawking thermal energy, $k_B T_H$, by the numerical dimensionless factor, $\eta = \pi\varepsilon(2\chi_{BH})^{3/2}$. Because $\varepsilon$ does not depend on parameters of the model $(\varepsilon \approx 0.16277)$, the numerical factor, $\eta$, depends, for a given axion mass, $m$, only on the total mass, $M$.

*Relation to the Unruh temperature.*

In our MF model, the gravitational interaction creates the acceleration for a single axion, due to its interaction with the rest of axions. As an example, consider the acceleration, $a$, at the characteristic radius, $r_{ch} \approx 8r_0$ (radius of external sphere in Fig. 8), of our MF system, where the ground state wave function vanishes,

$$a \approx \frac{GM}{r_{ch}^2}. \tag{5.6}$$

It is reasonable to ask a question: How the corresponding Unruh temperature [11-13],

$$T_U = \frac{\hbar a}{2\pi k_B c}, \tag{5.7}$$

is related to the MF ground state energy (5.1)? Using (2.10), (5.2)-(5.4), (5.6) and (5.7), we have from (5.1),

$$E \approx -128\pi\varepsilon\sqrt{\frac{2}{\chi_{BH}}} k_B T_U. \tag{5.8}$$

The dimensionless factor, $128\pi\varepsilon\sqrt{(2/\chi_{BH})}$, depends only on the total mass, $M$, for a given axion mass, $m$.

*Relation to the entropy of the BH.*

We will show that the entropy in our MF model $(S_{MF})$ is proportional to the square of the Schwarzschild radius. We can estimate the entropy in our model as,

$$S_{MF} = \ln(\Omega_N) = \ln(\Omega_1^N) = k_B N \ln(\Omega_1), \tag{5.9}$$

where $\Omega_1$ is the number of states for a single axion, and $N$ is the total number of axions, $N = M/m$.

Note, that in the MF limit, which we use, $\Omega_1 = 1$, so the entropy is zero. Let us assume that due to the interaction of axions with the environment, the value of $\Omega_1 > 1$, so the factor, $\ln(\Omega_1) > 0$. We have from the definition of $R_s$,

$$M = \frac{R_s c^2}{2G}. \tag{5.11}$$

Using (5.2), we have from (5.4),

$$m = \sqrt{2\chi_{BH}} \cdot \frac{\hbar}{R_s c}. \tag{5.12}$$

Using (5.11) and (5.12), we have for the total number of axions,

$$N = \frac{M}{m} = \frac{R_s^2 c^3}{2\sqrt{2\chi_{BH}} G\hbar}. \tag{5.13}$$

Finally, by substituting (5.13) into (5.9), we have for the dimensionless entropy in our model,

$$S_{MF} = \frac{\ln(\Omega_1)}{2\sqrt{2\chi_{BH}}\pi} \cdot S_{BH}, \left( \text{where, } S_{BH} = \frac{\pi k_B R_s^2 c^3}{G\hbar} \right). \tag{5.14}$$

We can see that the entropy, $S_{MF}$, in our model, differs from the entropy of the corresponding BH $(S_{BH})$ [10] by the numerical dimensionless factor, $\ln(\Omega_1)/(2\sqrt{2\chi_{BH}}\pi)$. In particular, the entropy, $S_{MF}$, is proportional to square of the Schwarzschild radius.

**Conclusion**

In this work, we have considered a DMH of a galaxy containing two species of ULA of different mass. Based on the mathematical theory [8], we derived the equations describing the ground state of the two-component BEC of ULA, in the MF limit. Generalizing a numerical method, developed in [7], we have studied numerically the radial distribution of the ULA mass density. We have constructed the parameters which can be used to distinguish a two-component DMH from the one-component DMH. We have also demonstrated an intriguing correlation between a one-component DMH in the MF limit and of a corresponding Black Hole temperature and entropy, and Unruh temperature.

**Acknowledgements**